\documentclass[english,twocolumn,prl]{revtex4-1}

\usepackage{graphicx}
\usepackage{amsmath}
\usepackage{amsfonts} 
\usepackage{bm}
\usepackage{color}
\usepackage{subfigure}
\usepackage{txfonts}
\usepackage{setspace}
\usepackage{ulem}

\begin{document}

\title{Encoding the structure of many-body localization with matrix product operators}

\author{David Pekker$^1$}
\author{Bryan K. Clark$^2$}
\affiliation{$^1$Department of Physics and Astronomy, University of Pittsburgh\\
$^2$Department of Physics, University of Illinois at Urbana Champaign}
\begin{abstract}
Anderson insulators are non-interacting disordered systems which have localized single particle eigenstates. The interacting analogue of Anderson insulators are the Many-Body Localized (MBL) phases. The natural language for representing the spectrum of the Anderson insulator is that of product states over the single-particle modes. We show that product states over Matrix Product Operators of small bond dimension is the corresponding natural language for describing the MBL phases. In this language all of the many-body eigenstates are encode by Matrix Product States (i.e. DMRG wave function) 
consisting of only two sets of low bond-dimension matrices per site: the $G_i$ matrix corresponding to the local ground state on site $i$ and the
$E_i$ matrix corresponding to the local excited state.  All $2^n$ eigenstates can be generated from all possible combinations of these
matrices.
\end{abstract}
\maketitle

An Anderson insulator is a non-interacting system that is driven into the insulating phase by quenched disorder~\cite{Anderson1958}. In one dimension, or for the case of strong disorder in higher dimensions, it is known that all single particle eigenstates can become localized~\cite{Abrahams1979}. When this occurs an Anderson insulator is a perfect insulator even at finite temperature. This should be contrasted with conventional insulators like band and Mott insulators that always display some form of activated behavior at finite temperatures.

The interacting analogue of the Anderson insulator is called many-body localization whose presence was originally suggested by a diagrammatic calculation~\cite{Basko2006, Basko2007}. Significant recent interest has gone to understanding whether many-body localized (MBL) phases exist as well as determining their properties~\cite{Oganesyan2007,Monthus2010,Huse2013,Serbyn2013,Huse2014,Agarwal2014}.  Many-body localized phases are believed to have a number of unusual properties including: (a) zero conductivity at finite temperature, (b) failure to thermalize, 
and (c) a large number of local constants of motion and corresponding conserved quantities. For MBL systems with a thermally-driven (or more precisely energy-density-driven) transitions these features persist all the way to the critical energy density.

The MBL phase transition is unique in that the phase transition is dynamical and, therefore, not simply a feature of the ground state wave-function or finite temperature density matrix. Instead, the MBL phase transition is believed to be caused by a qualitative change in the finite energy density eigenstates of the Hamiltonian.  In fact, it is known that MBL eigenstates are special with evidence accumulating that they obey an area law and exhibit poisson statistics. Understanding of these MBL eigenstates have come from exact-diagonalization studies of small (up to 16 sites~\cite{Pal2010,Bauer2013,Iyer2013,Kjall2014}), T-DMRG (time dependent density matrix renormalization group~\cite{Bardarson2012,Vasseur2014,BarLev2014}) and real space strong disorder renormalization group analysis~\cite{Vosk2013,Pekker2014}.

In contrast to the MBL eigenspectrum, the many-body eigenstates of a typical many-body Hamiltonian have essentially the same properties as arbitrary states sampled from the Hilbert space.  In fact, the eigenstate thermalization hypothesis (ETH) \cite{Deutsch1991,Srednicki1994,Rigol2008} suggests that typical eigenstates must locally look the same as the thermal density matrix at the temperature corresponding to their respective energy density. This means they obey a volume law: the entanglement entropy of a small subsystem with the remaining system is proportional to the volume of the subsystem. Moreover, level repulsion causes gaussian orthogonal ensemble level statistics in the eigenstates of a typical Hamiltonian. 

Like MBL eigenstates, the many-body eigenstates in an Anderson insulator have atypical properties.  In addition, though, they have a very simple form: a product state over localized single-particle eigenstates.  Importantly, this means that for an $L$-site lattice, $L$ single particle localized orbitals is sufficient knowledge to generate \emph{every} many-body eigenstate.  From this simple form, many of the properties of Anderson insulators can be understood. This leads us to a simple question: Do the many-body eigenstates of a MBL phase also share a simple and concise form? 

In this letter we use the language of matrix-product states to identify a simple form for MBL eigenstates.  To manifest this structure, we describe the case of an Anderson insulator and then show a natural generalization of the Anderson insulator case for the MBL case. Consider the specific example of one dimensional disordered spin-1/2 chains. In the non-interacting case, we can work in the basis of single particle eigenfunctions. Moreover, since all eigenfunctions are localized we can assign each eigenfunction to a lattice site $i$. Hence, each state of the many-body spectrum corresponds to a product state in which we assign each site of the lattice either $\{\psi_i=0, \phi_i=1\}$ if the corresponding single particle state is empty or $\{\psi_i=1, \phi_i=0\}$ if it is occupied
\begin{align}
\Psi_\text{Anderson}=\prod_i (\psi_i |e_i\rangle +\phi_i |g_i\rangle).
\end{align} 
A natural extension of these product states to the interacting but localized regime is obtained by replacing the the localized single particle orbitals $e_i$ and $g_i$ by tensors of finite bond dimension $E_{i,jk}^{\sigma_i}$ and $G_{i,jk}^{\sigma_i}$, where the indices $j$ and $k$ are dummy indices that are summed over when contracting the tensors. The index $\sigma_i$ corresponds to the local spin state on site $i$, which we will choose to be defined in the original Fock basis.  While these tensor states can encode some short distance entanglement, just like product states, they cannot encode long distance entanglement. We can then write
\begin{align}
\Psi_\text{MBL}=\prod_i  (\psi_i E_i^{\sigma_i} +\phi_i  G_i^{\sigma_i})|\sigma_i\rangle.
\label{eq:struct}
\end{align} 
Hence, the MBL ground state corresponds to the Matrix Product State (i.e. DMRG wave function)
\begin{align}
\Psi_\text{MBL,000\dots}=\sum_{jkl\dots} 
G_{1,j}^{\sigma_1} G_{2,jk}^{\sigma_2} G_{3,kl}^{\sigma_1}\dots |\sigma_1\sigma_2\sigma_3 \dots\rangle.
\end{align} 
Swapping $G_{i,jk}^{\sigma_i}$ for $E_{i,jk}^{\sigma_i}$ creates a local excitation of the system. 
\begin{align}
\Psi_\text{MBL,010\dots}=\sum_{jkl\dots} 
G_{1,j}^{\sigma_1} E_{2,jk}^{\sigma_2} G_{3,kl}^{\sigma_1}\dots |\sigma_1\sigma_2\sigma_3 \dots\rangle.
\end{align} 
The full many-body spectrum can be obtained by composing all combinations of $G_{i,jk}^{\sigma_i}$'s and $E_{i,jk}^{\sigma_i}$'s on all sites, thus mapping product states onto matrix product states. We show that these matrices can be directly identified from the Matrix Product Operator (MPO) derived from the unitary transformation that diagonalizes the Hamiltonian of our MBL system. Remarkably, we find strong numerical evidence that inside the MBL phases this MPO is efficiently representable: the typical bond dimension of the tensors $G_{i,jk}^{\sigma_i}$ and $E_{i,jk}^{\sigma_i}$ saturates at a finite value even as the system size becomes larger.

The paper is structured as follows. First, we give a brief description of MPS's and MPO's showing how an MPO can be used to represent the unitary transformation that diagonalizes the Hamiltonian.  This will imply that the eigenstates of the system can all be encoded in the structure of Eq.~\eqref{eq:struct}.  Then, we will argue that this MPO representation is compact -- the bond-dimension of a typical bond of the MPO is constant as a function of system size in the MBL phase.  To show this, we describe a numerical procedure for  constructing an MPO representation of the unitary transformation and apply this procedure to construct the MPOs for a large number of disorder strength, system sizes, and disorder realizations.  We devise an approach to match product states to eigenstates so as to minimize the `locality mismatch'.  From our numerical results we find that MPOs are indeed an efficient way to represent the many-body spectrum of an MBL system. At the same time we uncover strong Griffith effects that arise from rare regions of weak disorder. To conclude, we discuss how a description in terms of MPOs naturally leads to the conjectured properties of MBL phases, and comment on the limitations on using MPO representation for numerical diagonalization of strongly disordered Hamiltonians.

\begin{figure}
\includegraphics[width=0.9\columnwidth]{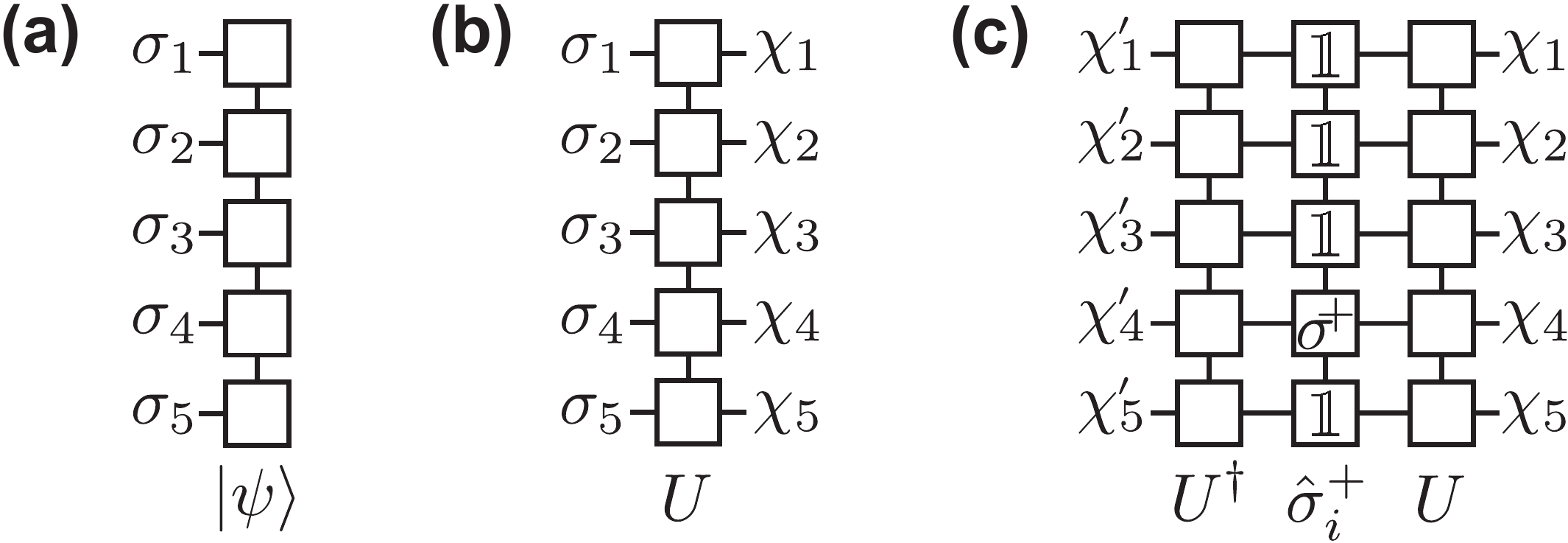}
\caption{
Box and line depiction of (a) a matrix product state $| \psi \rangle$, (b) a matrix product operator $U$, and (c) the product of three matrix product operators $U^\dagger \hat{\sigma}_i^+ U$. The boxes correspond to tensors and lines to tensor indices. Lines that connect two boxes correspond to indices that are contracted while dangling lines correspond to external indices. The matrix product state $| \psi \rangle$ ``eats" the external indices $| \sigma_1 \sigma_2 \sigma_3 \dots \rangle$ (these are the configurations of our spin-1/2 chain) and ``spits" out a complex number -- the amplitude of that configuration. Similarly, the matrix product operator ``eats" one set of external indices $| \sigma_1 \sigma_2 \sigma_3 \dots \rangle$ corresponding to the ``ket" and a second set of external indices $\langle \chi_1 \chi_2 \chi_3 \dots |$ corresponding to the ``bra" and ``spits" out the value of the corresponding matrix element.}
\label{fig:boxes}
\end{figure}

{\it An introduction to Matrix Product States and Operators} --  A MPS represents a quantum state and a MPO represents an operator. A convenient way to depict matrix product states and operators is using pictures composed of boxes and lines, see Fig.~\ref{fig:boxes}. For the case of spin-1/2 chains, the external indices can take on two values $\sigma_i, \chi_i \in \{|\uparrow\rangle_i, |\downarrow\rangle_i \}$. On the other hand the internal indices can span the range $\{1, \dots, D_{i}\}$, where $D_{i}$ is the ``bond dimension" for the bond between site $i$ and $i+1$. The value of $D_{i}$ is a tuning parameter that controls how much entanglement can be carried by the internal index linking neighboring sites. In practice, to describe eigenstates of strongly disordered systems we allow each internal bond to have a different bond dimension as dictated by the disorder realization. 

In summary, a MPS for an $L$-site chain is parametrized by $2L$ matrices -- two matrices per site $M_{i,k_i k_{i+1}}^\uparrow$ and $M_{i,k_i k_{i+1}}^\downarrow$. Analogously, an MPO contains four matrices for each site $i$: $O_{i,k_i k_{i+1}}^{\uparrow \uparrow}$, $O_{i,k_i k_{i+1}}^{\uparrow \downarrow}$, $O_{i,k_i k_{i+1}}^{\downarrow \uparrow}$, and $O_{i,k_i k_{i+1}}^{\downarrow \downarrow}$. 

We can represent the composition of operators and states in the matrix product language.  As an example consider performing a basis transformation on the single site spin raising operator $\hat{\sigma}_i^+$. $\hat{\sigma}_i^+$ can be represented as an MPO of $D=1$ with $O_j^{\alpha \beta}=\delta_{\alpha,\beta}$ if $j \neq i$ and $O_i^{\alpha \beta}=\sigma^+_{\alpha,\beta}$. If the unitary transformation is also represented as an MPO, we obtain a new MPO with $\tilde{O}_i^{\chi_i \chi'_i}=(U^*)^{\chi_i \alpha_i} O^{\alpha_i \alpha'_i}  U^{\alpha'_i\chi'_i}$ [see Fig.~\ref{fig:boxes}(c)].

When the unitary operator $U$, which diagonalizes a Hamiltonian, acts on one of the $2^L$ product states it returns the corresponding eigenstate of the Hamiltonian.   In the MPS language a product state is a MPS of $D=1$ and therefore when a product state is acted on by a MPO, the resulting MPS simply selects two of the four MPO matrices per site. Therefore, if we represent $U$ as an MPO, all the eigenstates of the Hamiltonian are encoded by matrix product states generated from all combinations of the matrices $O_i^{\downarrow \sigma_i}$ and $O_i^{\uparrow  \sigma_i}$. We then choose $G_i$ and $E_i$ to be $O_i^{\downarrow  \sigma_i}$ or $O_i^{\uparrow  \sigma_i}$ depending on whether the product state which maps to the ground state has down or up on site $i$. Notice that all eigenstates of the system are represented by $4L$ matrices (those that make up the MPO).  The key question, which we shall now address, is whether the MPO which represents the unitary operator can be represented by matrices with a finite bond dimension independent of the system length $L$.  

\begin{figure}
\includegraphics[width=\columnwidth]{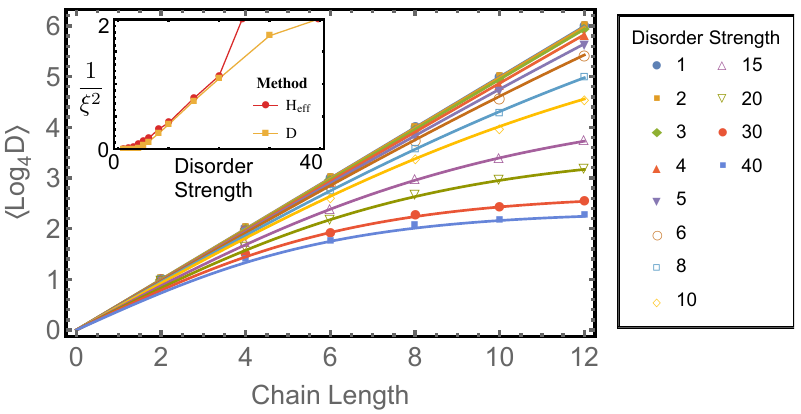}
\caption{
Bond Dimension of the Matrix Product Operator representing the unitary that diagonalizes the Hamiltonian as a function of the Chain Length for a number of values of disorder strength. The points represent numerical data for the bond dimension across the center bond of the chain, computed from an average of 200 realizations of disorder. The solid lines represent fits to the numerical data (see text). For systems with weak disorder the bond dimension grows linearly with the system size. On the other hand, for systems with strong disorder the bond dimension saturates at a finite value, even as the chain length keeps grows. The change of regime corresponds to the transition from the localized to the delocalized phase. Inset: Localization length $\xi^{-2}$ as a function of disorder strength. The localization length was computed in two different ways: (1) from the dependence of the coefficient of $J_{|i-j|} \sigma_i^z \sigma_j^z$ on the separation $|i-j|$ in the effective Hamiltonian~\cite{Huse2013a} and (2) from the fits to the saturation curves of the bond dimension in (d). In order to make the curves match, we had to rescale the localization length derived from bond dimension by a factor of $\sim8.7$.
}
\label{Fig:BDvsL}
\end{figure}

{\it Small bond dimension MPO} -- The strategy that we employ for testing whether MPOs can efficiently describe the unitary transformation that takes product states to eigenstates consists of three steps: (1) we construct the exact unitary transformation using exact diagonalization, (2) we identify a correspondence between the exact eigenstates and product states which maximally preserves locality, and (3) we compress the transformation into a matrix product operator. The Hamiltonian we test this approach on is 
\begin{align}
\label{eq:H1}
H=\sum_{\langle i,j\rangle} \sigma_i \cdot \sigma_j + \sum_i h_i \sigma_i^z
\end{align} 
where $h_i$ is a random field chosen from a distribution $h_i \in [-\Delta,\Delta]$. $H$ is known to have a MBL transition at $\Delta\sim3.5$~\cite{Pal2010}.  In our strategy, there is a clear notion of optimality: the procedure that produces an MPO with the smallest bond dimension should be considered optimal. This notion of optimality hinges on correctly identifying the spatial position of the excitations in the exact eigenstate and matching these locations to the product state. As we shall discuss in the next few paragraphs, finding the optimal MPO is a numerically challenging task. Therefore, we use a heuristic procedure to match eigenstates to product states, and so the bond dimension we obtain should be thought of as an upper bound to the best MPO bond dimension.

Our implementation of the exact diagonalization step is straightforward. The only point of note is that Hamiltonian~\eqref{eq:H1} conserves the total $S_z$, and therefore we diagonalize each $S_z$ subspace independently. Having obtained a list of eigenvectors we move onto the first of the two numerical challenges: how to relate the list of eigenvectors $P_E$ to the list of product states $P_P$.

The conjecture for why the unitary should be compressible into an MPO of low bond dimension is that all of the excitations of the system in the MBL phase must be spatially localized and the unitary therefore only needs to connect these spatially local sub-systems. This can be performed using a unitary with bond dimension $D \lesssim 4^l$, where $l$ is the characteristic lengthscale. However, to take advantage of this fact we must extract the locations of these localized excitations in each eigenvector and hence map the eigenvector onto the corresponding product state. 

In principle, we could try every possible match between the two lists $P_E$ and $P_P$ and select the one that produces the MPO with the lowest bond dimension; this procedure, however, is numerically intractable and therefore we use the  following heuristic approach. Consider the function $M(i)$ which specifies a unique matching of product states labelled
by $i$ to eigenstates $M(i)$.   Then define the objective function $\sum_i |\langle M(i)|i\rangle |^2$.   We  maximize
this objective function over functions $M$; this can be accomplished in polynomial time using the Hungarian 
algorithm for bipartite matching~\cite{Burkard1980}. 

The intuition for why this procedure is reasonable comes from thinking about MBL as Anderson localization in a many-body Hilbert space~\cite{Basko2006}.  We expect that eigenstates of a MBL phase have large IPR in the $\sigma^z_i$ basis (i.e. the eigenbasis of the non-interacting piece of the Hamiltonian \eqref{eq:H1}), and thus will have high overlap with the product states to which it should be mapped. We further optimize the matching task by matching only eigenstates to product states within the same total $\sigma^z$ sector.

Having performed the bipartite matching, we move on to the second numerical challenge: finding the MPO representation of our unitary transformation. Here, we can take advantage of the spatial information extracted in the previous step: we rewrite the unitary operator as a wave function by collapsing the two sets of external indices $\sigma_i$ for the eigenstate spins and $\chi_i$ for the product state spin  into a single index $\mu_i=2*\sigma_i+\chi_i$. This glues  together indices that are spatially close, meaning little entanglement exists between $\mu_i$ and $\mu_j$ if $i$ and $j$ are far apart. We then compress this state into a MPS.  Finally, we convert the MPS into an MPO by splitting the $\mu_i$ indices into the $\sigma_i$ and $\chi_i$ constituents. 

The bond dimension of the resulting MPO is set by a tuning knob  contained in the compression step where we select a cut-off for the smallest singular value we keep. If we select a lower cut-off, then we keep more singular values and hence obtain an MPO with a higher bond dimension which more faithfully matches the exact unitary.  Consider fixing the cut-off to a small value. If the unitary is indeed local, i.e. the Hamiltonian giving rise to it is many-body localized, we find that there are very few singular values above the cut-off and hence the resulting MPO has a small bond dimension. On the other hand if the unitary is non-local, i.e. the parent Hamiltonian is ergodic, there are many more singular values above the cut-off and hence the bond dimension of the resulting MPO is large. 

{\it Numerical Results}  -- The main result of our manuscript is depicted in Fig.~\ref{Fig:BDvsL}. In this figure, we plot the bond dimension $D$ of the Matrix Product Operator representing the unitary that diagonalizes the Hamiltonian \eqref{eq:H1} as a function of system size $L$ for various disorder strengths $\Delta$. In producing the plot, we have averaged the $\log_4[D]$ over many disorder realizations. As we are averaging the logarithm of the bond dimension, rare regions do not have a disproportionate affect on the average. From the figure, we observe that for systems with weak disorder ($\Delta \lesssim 3$) the bond dimension grows linearly with system size, while for those with strong disorder ($\Delta \gtrsim 15$) the bond dimension has saturated by the time the chain length has reached $L=12$. For disorder strengths $3 \lesssim \Delta \lesssim 15$ we do not have access to long enough chains to make a qualitative statement. 

We can, however, quantify the saturation effect by fitting the $D$ vs. $L$ curves with a generic saturation function: $\log_4[D(L)]=a \tanh(L/\xi)$ where $a$ and $\xi$ are the fitting parameters. In the inset of Fig.~\ref{Fig:BDvsL} we plot the saturation length scale $\xi$ as a function of the bond dimension. We observe that for systems with strong disorder the saturation length-scale is indeed very short. As the disorder becomes weaker, $\xi$ increases, becoming divergent for $\Delta \lesssim 3$. The $\xi$ we obtain matches that of Ref.~\cite{Huse2013a} [see caption Fig~\ref{Fig:BDvsL}].

\begin{figure*}
\includegraphics[width=\textwidth]{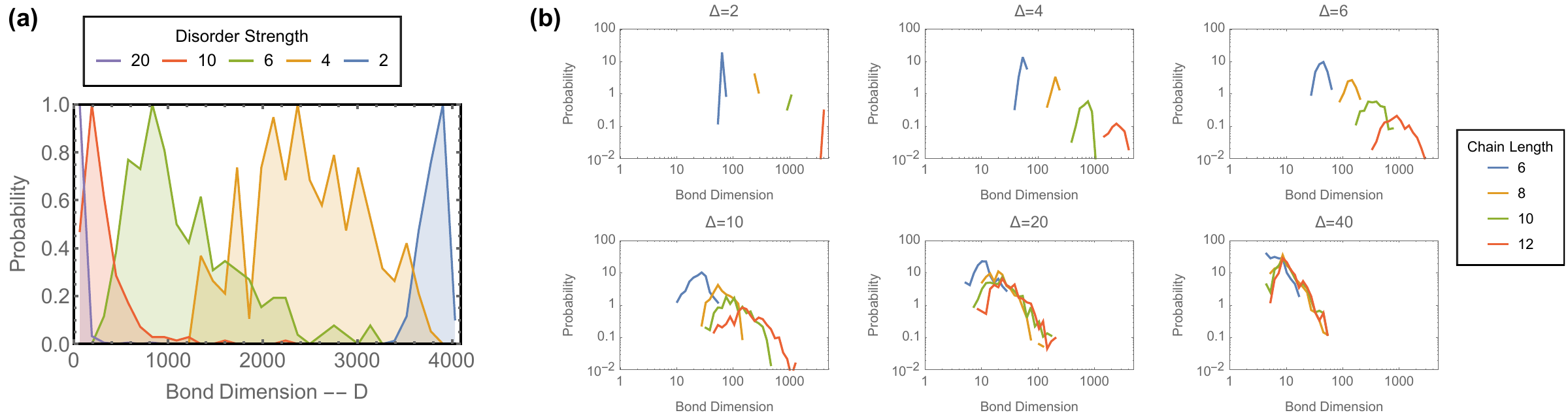}
\caption{
Probability Distribution Function (PDF) of the bond dimension across the center bond of the spin chain. (a) PDF for 12-site chains at various disorder strengths. For weak disorder, $\Delta=2$, the PDF is clustered around the maximal allowed value $D=4^6=4096$. Near the many-body localization-delocalization transition, $\Delta=4$, the PDF becomes spread over a wide range of bond dimensions. In the localized phase, the PDF becomes clustered around the $D=4^\xi$ with a power law tail extending to larger bond dimensions. (b) PDF for chains of various length and disorder strengths plotted on Log-Log axis. The formation of a Griffiths phase, in which the PDF has a power law form, can be clearly observed for $\Delta \gtrsim 6$. The effects of the chain lengths on the PDF become essentially invisible for $\Delta>20$ as the localization length becomes much shorter than the system size (although with better statistics we should be able to observe that the power law tails become cut off at $D\sim4^{L/2}$).  }
\label{fig:dist}
\end{figure*}

To summarize, our main result is that for systems in the MBL phase the full spectrum of eigenvectors in the localized phase can be described using an MPO of low bond dimensions. This observation dictates the structure of the many-body eigenstates. The properties of this structure can be used to identify the unusual properties, both static and dynamic, of MBL matter. Specifically, the MPO representation quantifies the notion of localized excitations and therefore dictates such properties as lack of thermalization, entanglement, emergent integrability, etc. We shall explicitly come back to these points in the discussion section, but first we explore how the MPO representation breaks down as we approach the delocalization critical point.

To understand the nature of the break down of the MPO representation, we look at the distribution of bond dimensions at fixed disorder strength. Specifically, we ask whether the distribution of bond dimensions is sharply peaked or not, and if not how does it behave. We summarize the behavior of bond dimensions in Fig.~\ref{fig:dist}(a). In the strongly localized matter, we find that the distribution of bond dimensions tends to be strongly peaked around $D=1$ (the minimum possible value for $D$). As the disorder strength decreases, we observe that (1) the peak in the distributions is starting to shift to small but finite values of $D$ associated with a growing localization length, and (2) the emergence of a power law tail in the distributions [see Fig.~\ref{fig:dist}(b)]. This power law tail signifies the onset of Griffiths physics: the system contains exponentially rare regions of the delocalized phase that give an exponentially strong contribution to the bond dimension~\cite{Bauer2013,Kjall2014,Agarwal2014}. As the disorder strength decreases the Griffith regions become less rare and begin to resonate. At the transition point we see a drastic change in the distribution of $D$ as it becomes extremely broad. This broadening, which has also been observed in the entanglement entropy of single eigenstates~\cite{Kjall2014}, culminates in the shift of the maximum of the distribution to a system size dependent value. The broadening of the distribution at the transition point indicates that the mechanism that drives the delocalization transition is the formation of resonances between the rare regions. 

{\it Discussion} -- The fact that the unitary that diagonalizes the Hamiltonian can be compressed into an MPO of small bond dimension has direct consequences for the properties of the MBL phase.  We begin by noting that the typical entanglement entropy of any of the eigenstates is finite as it is limited by $\log[D]$ which contradicts ETH.  Within our framework we can rule out thermalization without appealing to ETH. Consider a local operator such as $U \sigma^+_i U^\dagger$. Note this is the l-bit raising operator~\cite{Huse2013a} in the MPO language.  The application of the MPO composed from $UU^\dagger$ and the MPO composed from $U\sigma^+_iU^\dagger$ differ only on a single site [see Fig.~\ref{fig:boxes}(c)].  As the matrix on this site has a bond dimension which doesn't grow with system size, it will connect single eigenstates to a sub-extensive number of eigenstates all of which must have similar matrices far from the operator application.  This fact tells us that (1) there is no thermalization as a local kick to the system remains local; (2) there is no electrical conductivity as an excitation injected into the system at site $i$ remains put for very long times; and (3) there is no level repulsion as excitations from spatially distant operators $\sigma^+_i$ and $\sigma^+_j$ have no overlap. 

Finally, this MPO language lets us explicitly write the emergent local constants of motion.   A constant of motion is a hermitian operator which commutes with the Hamiltonian.  Consider operators of the form  
\begin{align}
\rho_\textrm{product}= I_1 \otimes \dots I_{k-1}  \otimes \sigma^z_k \otimes I_{k+1}  \dots I_{n} = \sum_p \alpha_p |p\rangle \langle p|
\end{align}
where $|p\rangle$ is a product state over all the sites. Applying the MPO $U$ to this operator gives us $U\rho_\textrm{product}U^\dagger = \sum_i \alpha_i |e_i \rangle \langle e_i|$ where $|e_i\rangle$ are eigenstates of the many-body system.  Operators of this form commute with the Hamiltonian and consequently 
are constants of motion.  Mirroring our previous argument, as $U\rho_\textrm{product}U^\dagger$ differ from $UU^\dagger$ by a single matrix they have an exponential weak effect on distant parts of the system and hence the constants of motion we've written down are local.

Finally we remark that the application of MPOs as a variational basis for diagonalizing many-body localized Hamiltonians has not escaped our notice. Indeed, our numerics indicates that in the localized phase we can represent the entire spectrum of  eigenstates of the Hamiltonian in a compact form using an MPO of low bond dimension. Due to the compact nature of the MPO representation it should be possible to diagonalize the Hamiltonian of rather large systems, significantly beyond the limits of exact diagonalization. The Griffiths effects will control the success of this endeavor. Specifically, each disorder realization will have rare regions of lower than typical disorder that will require an exponentially large bond dimension. The probability to find a rare region of length $l$ in a chain of length $L$ scales as $L \exp(-l/\xi)$. Therefore, with probability $1$ a chain will contain a rare region that requires $D \propto L^\xi$, which is a much softer constraint than the typical exponential scaling for exact diagon	alization. Although we save the construction of these MPOs for larger systems for a future work, we point out that having the complete spectrum will allow for efficient evaluation of finite energy density and dynamical properties of these systems.

In this work, we have focused on elucidating a structure for the entire spectrum of eigenstates that is analogous to the structure that is seen in Anderson localization.  We have additionally seen that the structure of these eigenstate gives us a very natural language to understand the property of the MBL phase.  Although we have focused here primarily on one-dimensional system, there is every reason to believe that the natural generalization where PEPS replace MPS will hold for higher dimensions.  

We thank Vadim Oganesyan, David Huse, Bela Bauer, and Chetan Nayak for useful discussions. We thank the KITP for its hospitality, DP acknowledges support from the Charles E. Kaufman Foundation and BK from grant DOE, SciDAC FG02-12ER46875.

Note added: during the preparation of this manuscript we became aware of a complementary work Ref.~\cite{Chandran2014}. 

\bibliography{dp}

\end{document}